\documentstyle[twocolumn,aps,prl,epsfig]{revtex}

\begin{document}
\draft
\preprint{HEP/123-qed}

\wideabs{
\title{\Large\bf  Structure of the most singular vortices
in fully developed turbulence}
\author{\normalsize{S. I. Vainshtein$^1$ and K.R. Sreenivasan$^2$}\\
{\small\it$^1$Department of Astronomy and Astrophysics, University
of Chicago, Chicago,  60637, USA\\
$^2$International Center for Theoretical Physics, Strada Costiera 11, I-34100
 Trieste, Italy}\\}
\date{\today}

\maketitle
\begin{abstract}
\normalsize{
Using high Reynolds number experimental data, we  search for most dissipative,
most intense
vortices. These structures possess a scaling
predicted by  log-Poisson model for the dissipation field
$\varepsilon_r$. 
These new experimental data suggest that the most intense structures have
co-dimension less than 2. The log-Poisson statistics is compared with
log-binomial which follows from the random $\beta$-model. 
}
\end{abstract}
}

\narrowtext

It is known, at least from numerical simulations, that the 
large-amplitude dissipation occurs around vortex tubes in turbulence. We thus expect some
structure to exist in a signal that characterizes the large values of the 
dissipation field. There are
some statistics, although very incomplete ones, on the distance between the
vortex tubes, the size of vortex tubes, etc.. The largest value of the 
dissipation is also important in determining the resolution of DNS \cite{sreeni}.
 It would be therefore of 
interest to provide a direct experimental study of the dissipative field extremal values.
On the other hand, very large values of the dissipation field correspond to intermittency.
Traditionally, the latter is expressed through so-called intermittency corrections to the 
exponents for the structure functions, $\langle |u(x+r)-u(x)|^p\rangle \sim 
r^{\zeta_p}$, where $u$ is the longitudinal velocity, and ${\zeta_p}=p/3$, \cite{k41}.
Thus, these corrections result in ${\zeta_p}=p/3+\tau(p/3)$. A
theory, incorporating the intermittency, the refined similarity 
hypothesis  \cite{kolm62}, links the
statistic of these corrections with the statistic of the dissipation field
$\varepsilon_r$, the energy dissipation averaged over a ball of size $r$. 
Namely,  $\langle \varepsilon_r^p\rangle \sim
r^{\tau(p)}$. Many models have been proposed to explain intermittency. It was
originally suggested that the statistics of  $\varepsilon_r$ is log-normal 
\cite{kolm62}. More
recently, She and L\'ev\^eque \cite{SL} (hereafter SL, see also \cite{D}, 
\cite{SW}, and recent study \cite{multiplier}) have proposed 
 log-Poisson statistics for the dissipation field, with agreement with 
the experimentally found $\zeta_p$ in \cite{Benzi},
\cite{Benzi1}. These experimental exponents are obtained in Extended
Self-Similarity approach, which is useful because of extended scaling range.

The simplest idea to study large values of dissipation is to measure the maxima. However,
for many distributions a maximum of a big array can be ``anything", or arbitrary large.
This is true, for example, for Gaussian statistics. The same is true for log-normal 
distribution.
 To see this, recall that, first, $\tau(p)=-d_p(p-1)$,
$d_p=D-D_p$, where $D_p$ are so-called generalized dimensions \cite{generalized}.
Second, studying maxima is in a way equivalent to measuring asymptotically high
moments, 
 $d_\infty=\lim_{p \to\infty}{\{-\tau(p)/(p-1)\}}$. For the log-normal distribution,
$\tau(p)=-\mu/2p(p-1)$, and therefore
asymptotically, $d_p=(\mu/2)p \to \infty$.
 Remarkably, the Poisson statistics
provide some distinctive maximum. To see this, recall that for the Poisson 
distribution (see, e.g., \cite{Poisson}), 
\begin{equation}
P(\alpha,\xi)=e^{-\xi} {\xi^\alpha}/{\alpha!},~~\alpha=0,1,2,...,~~ 
\varepsilon_r=e^{\alpha a+b},
\label{Poisson}
\end{equation}
and  $a<0$, the maximal value is defined through $b=\ln{\max{\varepsilon_r}}$.
In order to specify $b$, we calculate the moments, 
$\langle\varepsilon_r^p\rangle$. Noting that  $\langle\varepsilon_r\rangle=1$, we
get $b=\xi(1-e^a)$, where $\xi=C\ln{(\ell/r)}$, 
$\ell$ being external scale, and $C$ is a constant. As a result (of calculation of
the moments), we get,
\begin{equation}
\tau(p)=C[1-(1-\gamma/C)^p]-p\gamma,
\label{spectrum}
\end{equation}
where $\gamma=C(1-e^a)$ \cite{note}. Using (\ref{spectrum}), it is easy to show that 
this
time $d_\infty=\gamma$, which is a finite number.

SL is recovered from (\ref{spectrum}) if $C=2$,  $\gamma=2/3$, so that
$a=\ln{(2/3)}$, $C$ being the
co-dimension of most dissipative structures, and $\gamma$ is defined by the
dissipation rate, i.e., inverse time-scale, $1/t_r\sim r^{-2/3}$ \cite{SL}.

The meaning of $C$ becomes even more clear directly from (\ref{Poisson}): 
the most intense
fluctuations correspond to $\alpha=0$ (as $a<0$), so that the probability
$P(\alpha=0)=e^{-\xi}=(r/\ell)^C=
(r/\ell)^{(D-H_0)}$, $D$ is dimension of space ($=3$). Thus the Hausdorff 
dimension for most dissipative structures
in SL theory, $H_0=1$, i.e., the structures are filaments.
On the other hand, using  expressions for $b$, $\xi$ and
$\gamma$, we now  rewrite (\ref{Poisson}) as follows,
\begin{equation} 
\varepsilon_r =
e^{\alpha a}\max{\varepsilon_r}=e^{\alpha
a}\left(\frac{r}{\ell}\right)^{-\gamma}. 
\label{maxima}
\end{equation}
 Putting $\alpha=0$
in (\ref{maxima}), we can see that the most intense structures are expected to
scale $\sim r^{-\gamma}$. Thus, for the the log-Poisson statistics, the maxima of
$\varepsilon_r(x)$ are not ``anything", and they are supposed to be self-similar.

 This  scaling
is proved to be possible to verify experimentally.
We used 10 million points of atmospheric data, 
 with an estimated Taylor microscale Reynolds 
number 9540, (experiment A) and 40 million points for both longitudinal and transfer 
velocities (experiment B).
  The data are treated in  spirit of Taylor
hypothesis, that is, the time series is treated as one-dimensional cut of
the process. The dissipation rate can be written as
\begin{equation}
\varepsilon({\bf x})=\nu (\partial_i v_j \partial_i v_j+\partial_iv_j\partial_j v_i),
\label{eps}
\end{equation}
(summation over repeating induces). The second term on the rhs vanishes after averaging,
for homogeneous incompressible turbulence. The first term consists of 3 longitudinal
and 6 transverse components. Therefore, it is natural to present the dissipation as
\begin{equation}
\varepsilon_c(x)=\nu\left[3\left(\partial_x v_x(x)\right)^2+
6\left(\partial_x v_y(x)\right)^2\right].
\label{eps1}
\end{equation}
For isotropic turbulence, $\langle (\partial_x v_y(x))^2\rangle=
2\langle (\partial_x v_x(x))^2\rangle$, and therefore, following \cite{sreenietal},
we may consider three types of dissipation, 
longitudinal, transverse,
\begin{equation}
\varepsilon_l(x)=15\nu\left(\partial_x v_x(x)\right)^2,~~
\varepsilon_t(x)=(15/2)\nu\left(\partial_x v_y(x)\right)^2,
\label{eps2}
\end{equation}
and combined, (\ref{eps}).

\begin{figure}
\psfig{file=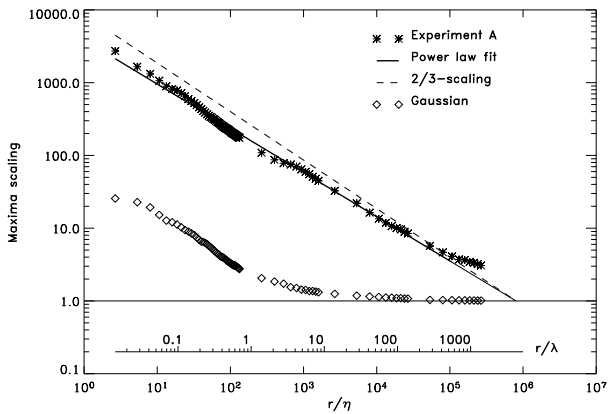,width=3.2in}
\caption{Scaling for most intense structures. The power law
fitting  of the experimental data (solid thick line) has been extended
to reach unity (solid  line), where it is supposed to match with SL
scaling. The distances are
given in terms of Kolmogorov micro-scale $\eta$, and in units of $\lambda$.}
\end{figure}

We will deal with coarse-grain dimensionless dissipation,
\begin{equation}
 \varepsilon_r=\frac{1}{r}\int_{x-r/2}^{x+r/2}\frac{\varepsilon(x')dx'}
{\langle \varepsilon \rangle},
\label{coarse}
\end{equation}
 and maxima of $\varepsilon_r$ can be measured. Note that this measurement is 
meaningful because (\ref{coarse}) contains some average. Figure 1
shows longitudinal scaling for the experiment A, which holds for $4.5$ decades. 
The deviation
from SL is small, and we recall that SL suggest that there is no anomalous
scaling for $t_r$. This small deviation in Fig. 1 can be interpreted as anomalous
persistence of the eddies, which is indeed observed \cite{persistence}, see also
discussion in \cite{our}.
The value of $\gamma$ is $0.61\pm 0.01$, only slightly smaller than $2/3$.
In order to compare with a ``regular" random process we generated a Gaussian
process $\omega_g$ with  correlation function coinciding with experimental,
i.e.,  $\langle \omega_g(x+r)\omega_g(x)\rangle=\langle
\partial_x v_x(x+r)\partial_x v_x(x)\rangle$. Then, the ``dissipation"
$\varepsilon^{(g)}=\omega_g^2$, and $\varepsilon^{(g)}_r=1/r
\int_{x-r/2}^{x+r/2}\varepsilon^{(g)}(x')dx'$. Corresponding calculation for 
the maxima are
reported in Fig. 1. If any scaling can be extracted from the Gaussian process, it
would be at large asymptotic distances, and the scaling is trivial, $\gamma=0$,
meaning no singularity. 

Figure 2 presents the scaling for experiment B. This time, the scaling holds for 
almost 6 
decades. The scaling exponent for longitudinal dissipation is again $0.61 \pm 0.01$, 
while for the combined dissipation (which is quite close to the transversal dissipation)
the exponent is $0.57 \pm 0.01$.
In both Figs. 1 and 2 there is also $\lambda$-scale. Note that there is characteristic
transfer region at  $r/\lambda \approx 1$. We may interpret it as a transition
to the inertial range \cite{vain}, which is formed due to the fact that the vortices
are expected to have scales between Kolmogorov microscale $\eta$ and Taylor microscale
$\lambda$.

\begin{figure}
\psfig{file=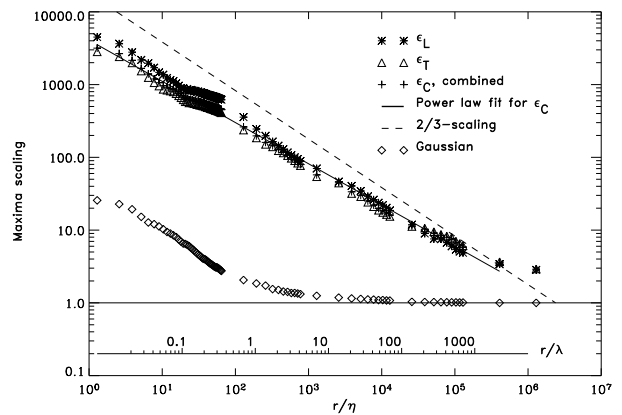,width=3.2in}
\caption{
Experiment B scaling for longitudinal and transversal dissipation, including combined
dissipation.}
\end{figure}

Although the experimental $\gamma$ is not that different from $2/3$, the value of
the other parameter $C$ is quite sensitive to that difference. 
In order to find $C$ we substitute  $\gamma$ from our measurements into
(\ref{spectrum}), and use computer routines to find a best fit for these data 
with free parameter $C$ and the exponents $\zeta_p^{(ESS)}$ from experiment 
\cite{Benzi}, \cite{Benzi1}. We start with the longitudinal dissipation (\ref{eps2})
from experiments A and B (recall that $\gamma$ is the same for them).
 As a result, we find
$C=1.67$ and $a=-0.45$ (cf. $\ln{\{2/3\}}=-0.41$). With these parameters, the
 deviation of these calculated exponents $\zeta_p^{(e)}$ from the experimental 
exponents, $\sqrt{\langle(\zeta_p^{(e)}-\zeta_p^{(ESS)})^2\rangle}=0.0063$. 
To compare: for SL, $\sqrt{\langle(\zeta_p^{(SL)}-\zeta_p^{(ESS)})^2\rangle}=
0.0078$. The  $\zeta_p^{(e)}$ exponents seem to be ``better" than 
$\zeta_p^{(SL)}$, but considering that the experimental exponents have  
errors of about $\pm 1\%$ \cite{Benzi1}, we conclude that these exponents are
similar.  
Note that if we substitute in
(\ref{spectrum}) the value of $\gamma=0.61$ and put $C=2$, then,
for the obtained exponents, $\zeta_p^{(C=2)}$, we have, 
$\sqrt{\langle(\zeta_p^{(C=2)}-\zeta_p^{(ESS)})^2\rangle}=
0.050$, much too high.

 Consider now the combined dissipation, defined in (\ref{eps1}).
The best fitting with this parameter fixed results in $C=1.43$, and $a=-0.50$. This
 time, the deviation of the computer generated spectrum from the experimental is
$=0.0071$.
 Figure 3 shows $\tau(p/3)$ from experiment, and for
different theories. It can be seen that all the curves collapse into one,
corresponding to the experiment, except that one with $\gamma$ from 
our measurements, and $C=2$. This illustrates that the data are indeed sensitive
to the measured $\gamma$, that is to its (small) difference from $2/3$.
 The codimension $C=1.43$ corresponds to $H_0=1.57$. This value of $H_0>1$ seems
to be consistent with the distinction between persistent vortical filaments and
the dissipative structures associated with regions of strong strain
\cite{vortex}.
That means that the most dissipative structures consist not only of 
filaments, but in part of sheets, or filaments convoluted into complex
structures, covering more than $1$ dimension. According to the
intersection theorem \cite{mandel}, that $D-H_0=D^{(m)}-H_0^{(m)}$,
where $D^{(m)}$ is the dimension of the measurements (in our case $D^{(m)}=1$),
and $H_0^{(m)}$ - corresponding measured Hausdorff dimension. It is clear from 
this formula that, if $H_0<2$ then $H_0^{(m)}<0$. This actually means that the
dimension $H_0 < 2$ cannot be detected in
1D measurements directly, and therefore our conclusion is inevitably indirect.
Indeed, it is obtained from spectrum (\ref{spectrum}), really formed in 3D, but 
projected into 1D assuming isotropy. Therefore,
it would be important to measure the Hausdorff dimension in 3D simulations
directly. Another reason for that is the surrogacy issue \cite{shreeni}.

These statements about the dimensions of the most
intense structures can be also formulated for log-binomial distribution, and, it
is known that the Poisson process is a limit of the binomial distribution for
``rare events". In particular, the Poisson distribution can be obtained
from the random $\beta$-model \cite{random} by a suitable limiting process 
\cite{D}, \cite{B}. Let $\beta$ take two values, $W=\beta_1$ with probability
$x$, and $W=\beta_2$ with probability $1-x$,
and $\beta_1x+\beta_2(1-x)=1$ (in order to have $\tau(1)=0$).
 Let also $\beta_1\leq 1\leq
\beta_2$.  Then, on the $n$-th level, the distribution is binomial, that is,
$W_n=\varepsilon_n=\beta_1^m \beta_2^{n-m}$ with probability
 $( ^n_m)x^m (1-x)^{n-m}$. Hence, $\langle \varepsilon_n^p\rangle=
[x\beta_1^p+(1-x)\beta_2^p]^n$. Taking into account that
$n=\ln{(r/\ell)}/\ln{\Gamma}$,
$\Gamma$ being the ratio of successive scales, we obtain,
\begin{equation}
\tau(p)=
\frac{\ln{[x\beta_1^p+(1-x)\beta_2^p]}}{\ln{\Gamma}}.
\label{spectrum1}
\end{equation}
In \cite{D} and \cite{B}, $\Gamma$ was treated as a free parameter. It was shown
that, if $\Gamma=1-x/C$,  $\beta_1=1-\gamma/C$ and $x\to 0$, then
$\beta_2\approx 1+x\gamma/C$, and
(\ref{spectrum1}) reduces to (\ref{spectrum}). 

The most intense structures on $n$-th level,
\begin{equation}
\beta_2^n=\left(\frac{r}{\ell}\right)^{\ln{\beta_2}/\ln{\Gamma}}=
\left(\frac{r}{\ell}\right)^{-\gamma_\beta}, 
\label{maxima1}
\end{equation}
cf. 
(\ref{maxima}). On the other hand, the probability of these maxima, 
\begin{equation}
P=(1-x)^n=\left(\frac{r}{\ell}\right)^{\ln{(1-x)}/\ln{\Gamma}}=
\left(\frac{r}{\ell}\right)^{C_\beta}.
\label{probability}
\end{equation}
In particular, if $\Gamma=1-x/2$ and $x\to 0$, then $C_\beta=2$ \cite{B1}.

If we do not treat  $\Gamma$ as a
free parameter, and consider that it is a fixed number, then 
the log-binomial 
distribution generally cannot be reduced to the log-Poisson PDF: in particular,
 if $x$ is small, then, $\tau(p)\sim x\to 0$, and thus
the intermittency is negligible . As in our case the division level $n\gg 1$, the
log-binomial distribution becomes essentially log-normal with maximum at $m=xn$
\cite{Poisson}, \cite{handbook}. However, the log-normal distribution has many 
shortcomings, and it has been repeatedly criticized when used to explain anomalous 
spectrum \cite{book}, \cite{Fbook}.
Nevertheless, the spectrum 
(\ref{spectrum1}) does not even look like log-normal
(for which $\tau(p)=-(\mu/2)p(p-1)$)
and rather behaves like log-Poisson for $p\gg 1$. Indeed, according to 
(\ref{spectrum1}), for $p\gg 1$,
\begin{equation}
\tau(p)=C_\beta +C_1\beta^p-p\gamma_\beta,~~
C_1=\frac{x}{(1-x)\ln{\Gamma}}<0,
\label{big}
\end{equation}
$\beta=\beta_1/\beta_2$.
This spectrum resembles (\ref{spectrum}); and the constants in (\ref{big})
happen to be numerically close to corresponding numbers in (\ref{spectrum}).
The reason for such a dramatic difference with log-normal distribution is as
follows. For binomial distribution,
\begin{equation}
\langle \varepsilon_n^p\rangle=\beta_2^{np}\sum_{m=0}^n  
(^n_m)x^m(1-x)^{n-m}\delta^m,
\label{sum}
\end{equation}
$\delta=\beta^p \ll 1$
 for large $p$. Then $\delta^m$ decreases dramatically with
increasing $m$, and therefore the terms of the sum (\ref{sum}) of maximal probability, at
$m\sim xn$, where  normal distribution if formed, do not
contribute substantially. In contrast, only the first few terms of this sum
(responsible for "rare" and very intense events) really contribute. 
Thus, effectively, the distribution works like a Poisson
distribution. To see this explicitly, consider a probability distribution
$(^n_m)x^m(1-x)^{n-m}\delta_0^m/A$, where $A$ is a normalization constant,
$A=(x\delta_0 +1-x)^n$, and $\delta_0=(\beta_1/\beta_2)^{p_0},~~ p_0\gg 1$.
Then, for large $n$ we express the factorials entering the binomial 
coefficients through Stirling formula (except for $m!$, because $m$ is not
necessarily large), to get,
\begin{equation}
P_1(m)=\frac{1}{A}(^n_m)x^m(1-x)^{n-m}\delta_0^m\approx e^{-\xi_0}
\frac{\xi_0^m}{m!},
\label{Poisson1}
\end{equation}
 where $\xi_0=n\delta_0 x/(1-x)$. For $p\ge p_0$, the sum (\ref{sum}) can
be written as $A\beta_2^{np}\sum_{m=0}^\infty P_1(m)\delta^{m(p-p_0)}$, and thus the
distribution  effectively corresponds to the  Poisson distribution.

Let us take the random $\beta$ model ``for real", that is, consider the Poisson
distribution as an approximation to the binomial, as in (\ref{Poisson1}).
 Then, we may consider that the model is realized as follows.
 Denote the number of divisions of each volume by $N$. Then $\Gamma=1/N^{1/D}$.
 We now multiply the values of $m$ divisions by $\beta_1$ and multiply the remaining
$N-m$ divisions by $\beta_2$. This actually means that the probability $x=m/N$, and
$1-x=(N-m)/N)$.  A particular
case of $N=2$ and $D=1$, i.e.,  $x=1/2$, corresponds to the model proposed in \cite{MS}. 
Then, according to
(\ref{probability}), $P=(r/\ell)^D$, i.e., the Hausdorff dimension $H_0$ is
$=0$, while $\gamma_\beta$, defined from (\ref{maxima1}), 
$=D\ln{\beta_2}/\ln{2}$. 
The case $\beta_1=0$ returns us to the $\beta$-model \cite{Frisch}. In that case,
$\beta_2=1/(1-x)$, and $\gamma_\beta=C_\beta=D\ln{(1-x)}/\ln{(1/2)}$. 

\begin{figure}
\psfig{file=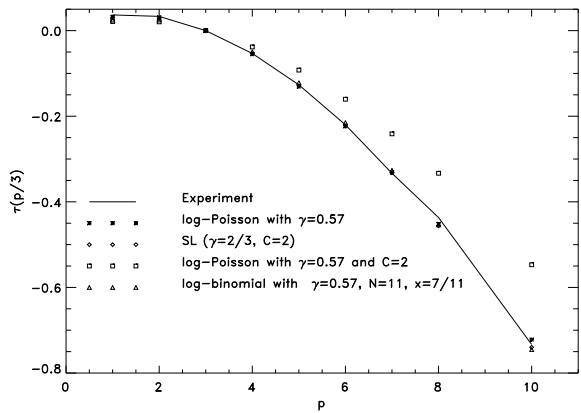,width=3.2in}
\caption{
Intermittency corrections from experiment  [\ref{Ben}],
 and from other 
theoretical models.}
\end{figure}

In general case, we may take $D=3$, and we are dealing with (\ref{spectrum1}) 
with $\gamma_\beta$ given from our measurements
(so that $\beta_2$ is defined according to (\ref{maxima1})).
We  thus are left with two free parameters, $N$, an integer, and $x=m/N$,
 where $m$ is also an integer. These numbers can be found with help of computer
search to fit experimental data \cite{Benzi}, \cite{Benzi1} in optimal way.
As a result of this search, we get: For $\gamma=0.57$ (combined longitudinal and
transversal dissipation), $N=11$, $x=7/11$. With these parameters, the deviation
of the spectrum from experimental is $0.0098$, quite satisfactory. 
 Indeed,
the corresponding $\tau(p/3)$  depicted in Fig. 3 is indistinguishable
from other approximations which collapse to the experimental data.
For $\gamma=0.61$ (longitudinal dissipation), $N=4$, $x=2/4=1/2$, $C_\beta=1.5$,
the deviation is $0.0115$, still okay. 
 As mentioned, at 
$p\gg 1$, the log-binomial spectrum (\ref{spectrum1}) is essentially reduced to the 
log-Poisson spectrum (\ref{spectrum}), and  therefore we would prefer to
consider the log-binomial distribution to be more general.

In conclusion, one of the predictions of SL theory about the scaling of maxima
$\sim r^{-\gamma}$ is experimentally confirmed. This makes it possible to make a 
better estimate of
the intense structures geometry in fully developed turbulence. The PDF's of the
exponents of the dissipation field are compared with the log-Poisson
distribution to show a good agreement with the theory. The log-Poisson statistic
can be considered as a limiting case for the log-binomial distribution appearing
in random $\beta$-model. We estimated the parameters of the log-binomial
distribution  with $\gamma$ found in our measurements, and to fit the exponents
for the structure functions found elsewhere. We conclude that the estimated
Hausdorff co-dimension of the most intense structures is less than 2.
 
 We appreciate numerous comments made by S. Boldyrev, Z.
Miki\'c, and R. Rosner.


\begin{references}
\bibitem{note}
It is easy to show that applying inverse Legendre transform to (\ref{spectrum})
we recover (\ref{Poisson}) if we express $\alpha!$ through the Stirling formula.
\end{references}
\end{document}